\documentclass[pra,preprint,aps]{revtex4-1}

\usepackage{graphicx}  
\graphicspath{{./Figures/}}
\usepackage{dcolumn}  
\usepackage{amssymb, amsmath}
\usepackage{natbib}
\usepackage{lineno} 
\usepackage{braket}
\usepackage{multirow}
\usepackage{bbold}

\begin{document}
\title{High-precision determination of the frequency-shift enhancement factor in Rb-$^{129}$Xe}
\author{A.I.\ Nahlawi}
\affiliation{Department of Physics and Astronomy, University of Utah, 115 South 1400 East, Salt Lake City, Utah, 84112-0830, USA}
\author{Z.L.\ Ma }
\affiliation{Department of Physics and Astronomy, University of Utah, 115 South 1400 East, Salt Lake City, Utah, 84112-0830, USA}
\altaffiliation[Present address: ]{Hill Air Force Base, Ogden, UT 63, USA}
\author{M.S.\ Conradi }
\affiliation{ABQMR, 2301 Yale Blvd S.E., Suite C-2, Albuquerque, NM 87106, USA}
\author{B.\ Saam}
\email{brian.saam@wsu.edu}
\affiliation{Department of Physics and Astronomy, Washington State University, P.O. Box 642814, Pullman, Washington, 99164-2814, USA}
\date{\today}
%
\begin{abstract}
We report a measurement of the dimensionless enhancement factor $\kappa_0$ for the Rb-$^{129}$Xe pair commonly used in spin exchange optical pumping (SEOP) to produce hyperpolarized $^{129}$Xe. $\kappa_0$ characterizes the amplification of the $^{129}$Xe magnetization contribution to the Rb electronic effective field, compared to the case of a uniform continuous medium in classical magnetostatics. The measurement is carried out in Rb vapor cells containing both $^3$He and $^{129}$Xe and relies on the previously measured value of $\kappa_0$ for the Rb-$^3$He pair. The measurement is based on (1) the optically detected (Faraday rotation) frequency shift of the $^{87}$Rb EPR hyperfine spectrum caused by the SEOP nuclear polarization and subsequent sudden destruction of nuclear polarization of both species and (2) a comparison of NMR signals for the two species acquired just prior to the EPR frequency shift measurements. We find $(\kappa_0)_{\rm RbXe} = 518\pm 8$, in good agreement with previous measurements and theoretical estimates but with improved precision.

\end{abstract}

\maketitle
%
\section{Introduction}\label{intro}
Almost six decades have passed since the first report of a hyperpolarized noble gas \cite{bouchiat1960nuclear}, but the field continues to grow and evolve, both in terms of the basic physics and the manifold applications. Nuclei of the stable spin-$\frac{1}{2}$ isotopes, $^3$He \cite{gentile2017optically} and $^{129}$Xe \cite{Meersman2014XeBook} may be polarized by spin-exchange optical pumping (SEOP) \cite{walker1997spin}, a two-stage process of angular momentum transfer. The ground-state electron spins of an alkali-metal vapor are polarized by the absorption of circularly polarized resonant light at the D$_1$ transition (${\rm 5S_{1/2}\rightarrow 5P_{1/2}}$ in Rb); subsequent collisions of the polarized alkali-metal atoms with noble gas atoms mediate an interatomic Fermi-contact interaction through which electron and nuclear spin is exchanged. The result is an ensemble of noble-gas atoms with a nuclear-spin polarization in the range of 10-100\%, several orders of magnitude beyond the thermal-equilibrium value at room temperature in even the largest laboratory magnetic fields. Hyperpolarized noble gases (in some cases along with a cohabitating polarized alkali-metal vapor) are used in sensitive magnetometry \cite{Limes2018}, inertial guidance \cite{WalkerBookCh2016}, and the search for physics beyond the Standard Model \cite{Chupp2019RMP,Almasi2018,Kuchler2016,Allmendinger2014}. They are additionally used as a sensitive signal source in magnetic resonance imaging \cite{Mugler2013Hyperpolarized1M,Leawoods2001}.

EPR of the alkali-metal hyperfine structure can be a sensitive embedded probe of the magnetic field generated by the evolving noble-gas nuclear magnetization in a SEOP cell. The same Fermi-contact interaction that transfers spin to the noble gas also produces a shift in the alkali-metal EPR frequency $\nu_A$ that is proportional to the nuclear magnetization \cite{schaefer1989frequency}:

\begin{equation}
\label{eq:EPRshift}
\Delta|\nu_{\rm A}|=\frac{1}{h}\frac{8\pi}{3}\Bigg|\frac{d\nu_{\rm A}}{dB_0}\Bigg|M_{\rm X}\kappa_{\rm AX},
\end{equation}

\noindent where the nuclear magnetization $M_{\rm X}$ is given by:

\begin{equation}
\label{eq:magdef}
M_{\rm X} = \mu_{\rm X}\frac{\langle K_z\rangle}{K}[{\rm X}].
\end{equation}

\noindent In Eqs.~(\ref{eq:EPRshift}) and (\ref{eq:magdef}), $[{\rm X}]$, $\mu_{\rm X}$, and $K$ are the noble-gas number density, magnetic moment, and spin, respectively; $h$ is Planck's constant, $B_0$ is the applied magnetic field, and $\kappa_{\rm AX} > 0$ is a dimensionless factor specific to each alkali-metal/noble-gas pair that parameterizes the enhancement of the noble-gas magnetic field sensed by the alkali-metal electrons. The enhanced field results from the quantum mechanical overlap of the electron wave function at the noble-gas nucleus, time-averaged over many collisions. A value of unity for $\kappa_{\rm AX}$ corresponds to the hypothetical case where the electron classically overlaps a continuous uniform noble-gas magnetization having the same value as that calculated by Eq.~(\ref{eq:magdef}) for discrete noble-gas atoms. The enhancement is $\approx 5$ in the case of Rb-$^3$He \cite{romalis1998accurate} and about 500 in the case Rb-$^{129}$Xe \cite{ma2011collisional}, owing to xenon's much greater atomic number.

In this paper we present a precise measurement of $\kappa_0$ for the Rb-$^{129}$Xe pair; the limit $\kappa_{\rm AX}\rightarrow \kappa_0$ obtains for sufficiently high third-body gas pressure (see Sec.\ \ref{theory} below). Our measurement is based on a ratiometric comparison of the optically detected $^{87}$Rb EPR frequency shifts due to $^{129}$Xe and $^3$He. Our result, $(\kappa_0)_{\rm RbXe} = 518 \pm 8$, is in good agreement with a previous measurement, $493 \pm 31$, based on the measured ratio of $^{129}$Xe and $^3$He NMR frequency shifts \cite{ma2011collisional}. It is also in good agreement with a recent theoretical prediction, $588 \pm 50$, based on detailed electronic-structure calculations \cite{hanni2017electron}. A precise measurement of $(\kappa_{\rm AX})_{\rm RbXe}$ is relevant to better understanding of SEOP physics for the Rb-Xe pair \cite{Korver2015} and vital for understanding and correcting systematic shifts in several tests of fundamental symmetries that feature alkali-metal and/or noble-gas magnetometers and co-magnetometers \cite{allred2002high}. It can also be used to calibrate {\it in situ} polarimetry of hyperpolarized $^{129}$Xe as it is produced by SEOP for the various applications.

\section{Theory}\label{theory}
For an applied magnetic field $B_0$ where the alkali-metal hyperfine splitting is large compared to the electron Zeeman splitting, the effective alkali-metal gyromagnetic ratio for $\Delta m_F=\pm 1$ transitions within the same hyperfine manifold, correct to linear terms in $B_0$ is \cite{breit1931measurement}

\begin{equation}
\label{eq:alkaligyro}
\Bigg|\frac{d\nu_{\rm A}}{dB_0}\Bigg|_{I\pm 1/2}=\frac{|g_s|\mu_B}{2I+1}\Bigg(1\mp \frac{2\overline{m}_F|g_s|\mu_B}{A(2I+1)}B_0+\mathcal{O}({B_0}^2)+ \cdots\Bigg) ,
\end{equation}
\noindent where $I$ is the alkali-metal nuclear spin, $g_s = -2.0023$ is the free-electron g-factor, and $A$ is the alkali-metal hyperfine coupling strength. The total angular momentum quantum number is $F=I\pm 1/2$, corresponding to the two hyperfine manifolds; and $\overline{m}_F$ is the mean value of the two magnetic quantum numbers for the neighboring levels involved in the transition. Eq.~(\ref{eq:alkaligyro}) represents the quadratic Zeeman splitting into $4I$ hyperfine spectral lines, one for each $F,\overline{m}_F$ pair. Under our experimental conditions, a highly polarized vapor means that Rb atoms are pumped into one of the end states $m_F = \pm F$ of the $I+1/2$ manifold by $\sigma^{\pm}$ polarized light.

Spin exchange is mediated through both binary Rb-Xe collisions and the formation of longer-lived RbXe van der Waals molecules \cite{happer1984polarization}. As described in detail by Schaefer, et al.~\cite{schaefer1989frequency}, the enhancement factor may be written as:

\begin{equation}
\label{eq:kappaAX}
\kappa_{\rm AX}=(\kappa_0-\kappa_1)+\epsilon_{\rm AX}\kappa_1,
\end{equation}

\noindent where $0\leq\epsilon_{\rm AX}\leq 1$ characterizes the fractional suppression of the enhancement that occurs as the mean lifetime of RbXe van der Waals molecules increases, i.e., as the mean precession angle $\phi$ of the coupled angular momenta about the molecular magnetic field during a RbXe van der Waals molecular lifetime approaches and exceeds one radian \cite{Zeng1985}. At sufficiently large third-body pressure (short molecular lifetime), $\epsilon_{\rm AX}\rightarrow 1$ and $\kappa_{\rm AX}\rightarrow \kappa_0$. According to Zeng, et al.\ \cite{Zeng1985},

\begin{equation}
\label{eq:phi}
\phi=\frac{p_0}{p},
\end{equation}

\noindent where $p$ is the total gas pressure and $p_0$ is a pressure that depends on gas composition and characterizes the transition from short to long molecular lifetime. For the Cs-Xe pair, $p_0 = 384$~torr if the third-body gas is He \cite{Hsu1985}; we assume that this number is not very different for the Rb-Xe pair. (For N$_2$ as the third body, the difference is about 30\% \cite{schaefer1989frequency}.) All of our cells contain at least four times this characteristic pressure of He. From Tab.\ II in Schaefer, et al.\ \cite{schaefer1989frequency}, the ratio $\kappa_1/\kappa_0\approx 0.08$ for the Rb-Xe pair; from Eq.~8 in Ref.\ \cite{schaefer1989frequency}, we calculate $\epsilon_{\rm AX} \geq 0.995$ for all of our cells. We conclude that for our experiments the true value of $\kappa_{\rm AX}$ deviates from $\kappa_0$ by less than 0.1\% in all cases, and we treat the effect of $\kappa_1$ as negligible.


In Rb vapor cells containing both $^3$He and $^{129}$Xe, one can perform rapid consecutive measurements of the EPR frequency shifts due to each noble-gas species (see Sec.\ \ref{expt}). From Eq.~(\ref{eq:EPRshift}) the ratio of these shifts is

\begin{equation}
\label{eq:ratio}
\frac{\Delta\nu_{\rm Xe}}{\Delta\nu_{\rm He}}=\frac{(\kappa_0)_{\rm RbXe}}{(\kappa_0)_{\rm RbHe}}\frac{M_{\rm Xe}}{M_{\rm He}}.
\end{equation}

\noindent The ratio $M_{\rm Xe}/M_{\rm He}$ can be determined by NMR measurements of free-induction decay (FID) signals for the two species.  Using the previously measured value of $(\kappa_0)_{\rm RbHe}= 4.52 + 0.00934T$, where $T$ is the temperature in $^{\circ}$C \cite{romalis1998accurate}, Eq.~(\ref{eq:ratio}) can then be used to determine $(\kappa_0)_{\rm RbXe}$. Both the EPR frequency shift and the NMR signal intensity depend only on the product of the nuclear polarization and the noble-gas density, i.e., the magnetization $M$ in Eq.~(\ref{eq:magdef}); therefore, the measurement of $(\kappa_0)_{\rm RbXe}$ requires no knowledge of the polarization of either noble-gas species and is insensitive to uncertainties in the cell pressures listed in Table I. The accuracy of the measurement comes down to (1) the degree to which we ensure the same initial nuclear magnetization for the EPR and NMR measurements (separately for both $^3$He and $^{129}$Xe), and (2) the degree to which the recorded measurements of the EPR and NMR signals are proportional to those magnetizations.

\section{Experiment}\label{expt}
We performed experiments on three spherical vapor cells of similar volume made of borosilicate (Pyrex) glass; they were filled with several tens of Torr of N$_2$ and Xe gases (the Xe is enriched to 90\% $^{129}$Xe), about 2000 Torr $^3$He and a few milligrams of Rb (see Table I). The oven temperature for SEOP ranged between 125~$^{\circ}$C and 175~$^{\circ}$C, as determined by a resistive thermometric device (RTD) affixed directly to the cell, although laser heating likely still caused the actual interior cell temperature to be somewhat warmer. In general, both $^3$He and $^{129}$Xe were polarized to their respective maximum (saturation) values; we note that this typically took many hours for $^3$He and minutes for $^{129}$Xe, and that the latter could be polarized at lower temperatures to its saturation value several times in one experiment with little to no effect on the $^3$He polarization. In all cases, $^3$He and $^{129}$Xe were polarized into the low-energy Zeeman state using $\sigma^-$ light. The SEOP pump laser was a 30-watt diode-laser array model A317B (QPC Lasers), externally tuned to the 795 nm D$_1$ resonance and narrowed to $\approx 0.3$~nm with a Littrow cavity \cite{chann2000frequency}. This was an older laser with at least many hundreds of hours of use, and the total output power had deteriorated somewhat; the narrowed output incident on the cell was typically 10-12~W.

Two distinct types of measurements were made: (1) Separate measurements of the $^3$He and $^{129}$Xe NMR signal intensities acquired with a Redstone NMR spectrometer (Tecmag); these measurements were made sequentially with a known RF-excitation (flip) angle at one Larmor frequency, adjusting the applied magnetic field $B_0$ to the corresponding value for each nucleus (see Sec.\ \ref{flipangle} below); typical data is shown in Fig.~1. (2) Sequential optically-detected (Faraday rotation) EPR measurements of the $^{87}$Rb frequency shift made at the same stable value of applied magnetic field (26.5~G); one made after destruction of the $^3$He magnetization and one made after destruction of the $^{129}$Xe magnetization; the destruction was accomplished with several near-90$^{\circ}$ resonant RF pulses from the Redstone spectrometer, which was switched between the respective Larmor frequencies; typical data is shown in Fig.~2. Fig.~3 is a schematic diagram for the entire experiment, showing alignments of the pump and probe lasers, main-field and excitation coils, and the detection and amplification chains for both NMR and EPR.

\subsection{NMR}\label{NMR}
The NMR probe was a parallel-capacitance-tuned pair of coils (3~cm diam) separated by about 5~cm, each with multiple layered windings of 25/45 Litz wire (coil inductance was $\approx 30~\mu$H). Since the goal was to measure the signal ratio for $^{129}$Xe and $^3$He at known flip angles for each nucleus, we sought to eliminate all other sources of instrumental discrepancies and systematic effects by using the same cell, along with the same NMR spectrometer, frequency, probe and amplifier settings for both measurements. For our operating Larmor frequency of 31.25~kHz, this required adjusting $B_0$ to 26.5~G for $^{129}$Xe and 9.6~G for $^3$He. A home-built pre-amplifier coupled the probe to the spectrometer; signals were displayed and analyzed using the NTNMR (Tecmag) software. The spectrometer was also used (in transmit mode only) for rapid destruction of $^{129}$Xe and $^3$He polarizations at the frequencies 31.25~kHz and 85.6~kHz, respectively, as further discussed in the next section.

\subsection{Optically Detected EPR}\label{EPR}
EPR of $^{87}$Rb was optically detected in a manner similar to that described by Chann, et al.\ \cite{Chann2002}, and a double-homodyning scheme was used to lock the EPR frequency to the measured magnetic field.  A 80-mW external-cavity-tuned probe laser, model DL-7140-201S (Sanyo), was directed transverse to the main applied magnetic field through the cell and then detected by a fast (2-ns rise time) silicon PIN photodiode, model 54-520 (Edmund Optics). The probe laser was detuned $\approx 1$~nm from the $^{87}$Rb D$_2$ resonance at 780~nm, where Faraday rotation of the plane of polarization has a well-characterized linear dependence on the $^{87}$Rb spin polarization along the propagation axis of the probe beam. The probe-beam diameter is a few millimeters, small compared to the cell diameter. A weak cw RF excitation generated a steady-state precessing transverse $^{87}$Rb magnetization at frequencies corresponding to the hyperfine resonances ($\approx 19$~MHz at 26.5~G for $^{87}$Rb). The resonance intensity at each hyperfine frequency was measured by converting the light-polarization modulation to a light-intensity modulation: a linear polarizer oriented at 45$^{\circ}$ to the nominal plane polarization of the probe laser was placed in front of the photodiode detector. The detector output was homodyned with the RF source, a voltage-controlled oscillator (VCO), model 80 (Wavetek). The VCO could be swept with a voltage ramp across all of the resonances to generate the full hyperfine spectrum, where we note that the spectrum under optical pumping conditions is asymmetric: most of the intensity resides in the $F=2$ end resonance that corresponds to the helicity of the D$_1$ pumping light; see Fig.~4. Operating the VCO with a low-frequency ($\approx 100$~Hz) modulation having an amplitude small compared to the resonance width generated the derivative of the more intense end resonance; the homodyned photodetector output was fed to a lock-in amplifier (model 186A; Princeton Applied Research) referenced to the modulation frequency; see Fig.~3. The zero-crossing of the derivative signal at the resonance peak was used in a loop fed back to the VCO to lock the EPR frequency, which was read out on a precision frequency counter, model 53220A (Agilent). In an otherwise stable applied field, the locking circuit could reproducibly follow both gradual and sudden changes in the noble-gas nuclear magnetization.

\subsection{Stabilization of applied magnetic field}\label{stabilize}
The measured EPR frequency shifts due to nuclear polarization were 1-10~kHz; we thus needed to stabilize the applied magnetic field $B_0$ to much better than a part in $10^4$. This was accomplished with the home-built current-stabilization circuit designed by one of us (M.S.C.) and shown in Fig.~5. Most commercial DC power supplies have some difficulty driving inductive loads in current-control mode. The supply used here (model 6267B; Hewlett-Packard) was run in voltage-control mode in series with a power MOSFET and a 250~W 0.5~$\Omega$ resistor (Vishay-Dale) having a temperature coefficient of 10 ppm/$^{\circ}$C. The resistor was immersed in mineral oil for better thermal stability. The current through the MOSFET, running well out on the flat portion of the $I$-$V_{\rm DS}$ curve, was controlled via the gate input generated by comparing the voltage across the 0.5~$\Omega$ resistor with a stable voltage reference, thus stabilizing the current through the Helmholtz coils. The Helmholtz-coil current was additionally monitored with a precision digital multimeter (model 2000, Keithley) measuring the voltage across a 0.10~$\Omega$ stable resistor in series with the coils only. The monitored stabilized current was used to correct for baseline drifts in frequency-shift data acquired over long time periods (many tens of minutes to hours). By running at certain times in the day (usually in the middle of the night) and employing these field-stabilization and monitoring schemes, the rms frequency noise (typically integrated over 1.5 seconds) was reduced to $\leq 100$~Hz, or about one part in $2\times 10^5$ of the EPR frequency. Data were acquired and stored with the help of LabVIEW software (National Instruments) on a PC-type computer.

\section{Results and Analysis}\label{results}
An experimental run consisted of (1) a NMR flip-angle measurement, (2) a measurement of the relative intensity of $^{129}$Xe and $^3$He NMR signals, and (3) a measurement of the relative size of the corresponding $^{87}$Rb EPR resonance shifts due to the $^{129}$Xe and $^3$He magnetizations.

\subsection{Flip-angle measurement}\label{flipangle}
At some time after both $^3$He and $^{129}$Xe polarizations had reached their maximum (saturation) values ($>20$~h of SEOP), we started an experimental run by measuring the magnetization ratio with NMR at 31.25~kHz. The measured initial height $S$ of the free-induction decay (FID) is proportional to the longitudinal magnetization from which it was generated through excitation at some flip angle $\theta$. The magnetization ratio just prior to excitation is given by:

\begin{equation}
\label{eq:signal}
\frac{M_{\rm Xe}}{M_{\rm He}}=\frac{S_{\rm Xe}}{S_{\rm He}}\frac{\sin\theta_{\rm He}}{\sin\theta_{\rm Xe}}.
\end{equation}

\noindent Had we been able to operate in the regime where $\theta \ll 1$~radian for both nuclei, then the sine ratio in Eq.~(\ref{eq:signal}) could be replaced by a factor of the ratio of gyromagnetic ratios ${\gamma_{\rm He}}/{\gamma_{\rm Xe}}$~\cite{saam1998low}, but this was not the case, and we had to carry out frequent and precise measurements of the flip angles used. This was done in a separate measurement by polarizing the $^3$He to saturation and then applying a long series of identical 80-$\mu$s pulses at 31.25~kHz, spaced 55~ms apart. The FID acquired after each pulse was Fourier transformed and the area under the peak was measured; see Fig.~6. The magnetization after the $n^{th}$ pulse was

\begin{equation}
\label{eq:decay}
M(n) = M_0\cos^n\theta = M_0e^{n\ln(\cos\theta)},
\end{equation}

Defining $b\equiv \ln(\cos\theta)$, we fit data such as that shown in Fig.~6 to the function $ae^{bn}$ with fitting parameters $a$ and $b$ and then calculated

\begin{equation}
\label{eq:thetab}
\theta=\arccos(e^{b}).
\end{equation}

Provided we used all of the same electronics and the exact same gains and other settings, the $^{129}$Xe flip angle was found from

\begin{equation}
\label{eq:HeXe}
\theta_{\rm Xe}=\frac{\gamma_{\rm Xe}}{\gamma_{\rm He}}\theta_{\rm He}.
\end{equation}

\noindent The output of the NMR pulse amplifier was monitored with an oscilloscope before and after every NMR measurement to ensure that the pulse characteristics in the flip-angle measurement remained consistent over the course of the entire experimental run.

\subsection{NMR signal acquisition}\label{NMRacq}
The $^{129}$Xe NMR signal in our cells, particularly at low flip angles, is quite weak but recovers in minutes with SEOP. The $^3$He signal is strong but needs tens of hours to recover if destroyed. The EPR frequency shifts from $^3$He are also smaller than those for $^{129}$Xe. We needed to choose a $^3$He flip angle low enough to record the NMR signal and still have plenty of magnetization left to record a significant EPR frequency shift when the remaining magnetization was destroyed. If we wanted to use the exact same pulse (frequency, power, and duration) for both species, we thus needed to acquire the $^{129}$Xe FID at an even smaller flip angle and then average many such acquisitions.

Starting at full polarization for both $^3$He and $^{129}$Xe and with the SEOP pump laser on continuously, we applied $N=100$ pulses 50~ms apart at the $^{129}$Xe Larmor frequency of 31.25~kHz in a field $B_0=26.5$~G. The acquired FIDs were added together and Fourier transformed. We assumed
that there was no significant additional polarization created by SEOP during the total acquisition time of 5~s; see Sec.\ \ref{discussion}. Due to the magnetization destroyed after each pulse, the total measured signal $S_N = \Sigma_{n=1}^{N}S_n$ needed correction to represent the signal $S_{\rm Xe}$ after the first measurement:

\begin{equation}
\label{eq:Sonebar}
S_{\rm Xe} = \frac{S_N}{\Sigma_{n=0}^{N-1}\cos^n\theta}.
\end{equation}

Immediately after this series of $^{129}$Xe NMR signal acquisitions,
the applied field was lowered to 9.6 G by reducing the current in the power supply, corresponding to a $^3$He Larmor frequency of 31.25~kHz. A single 80-$\mu$s excitation pulse was applied, and $S_{\rm He}$ was immediately recorded for $^3$He without signal averaging. The pulse length, frequency, and power were unchanged between the $^3$He and $^{129}$Xe NMR measurements, as were the amplifier gain settings on the receiver side.

\subsection{EPR frequency shift acquisition}\label{EPRacq}
After the NMR measurements, the $^{129}$Xe magnetization was allowed to recover by SEOP (typically requiring no more than a few minutes). The applied magnetic field was returned to 26.5~G and stabilized using optically detected $^{87}$Rb EPR, as described in Sec.\ \ref{EPR}. A baseline EPR frequency was established prior to destroying the remaining $^3$He magnetization with a rapid series of large-angle pulses at 85.6~kHz. A new baseline was then established and the total shift $\Delta\nu^{\prime}_{\rm He}$ was recorded. We use the prime because this shift still had to be corrected for the magnetization lost from the one pulse used previously to acquire a single FID:

\begin{equation}
\label{eq:delnu}
\Delta\nu_{\rm He}=\frac{\Delta\nu^{\prime}_{\rm He}}{\cos\theta_{\rm He}}.
\end{equation}

\noindent The $^3$He magnetization, once destroyed, did not recover to any significant degree during the remainder of the experimental run due to the low spin-exchange rate; this was verified at the very end of the run by repeating the above steps and noting no significant frequency shift.

A few minutes after the $^3$He measurement, the $^{129}$Xe polarization was similarly destroyed with a series of pulses at 31.25~kHz. Because the $^{129}$Xe magnetization recovered quickly via SEOP (within 3~min), we repeat its destruction and recovery five times for each experimental run, averaging the results to obtain the $^{129}$Xe frequency shift $\Delta\nu_{\rm Xe}$; this shift requires no correction, since the starting point is full polarization prior to each destructive series of pulses; see Fig.~2.

There were a total of 48 experimental runs with three different sample cells (Table I) and four different temperatures. The value of $\kappa_0$ for RbXe and associated uncertainty was calculated using Eq.~(\ref{eq:ratio}) for each run; results are shown in the plot in Fig.~7. Table II shows the uncertainty-weighted average values categorized according to both temperature and the sample cell used. The weighted average value for all measurements yields $(\kappa_0)_{\rm RbXe}= 518\pm 8$ (green line in Fig.~7), where 8 is the standard deviation of the mean (dark green region) and the light green region ($\pm 59$) corresponds to the sample standard deviation.

\section{Discussion and Conclusion}\label{discussion}
There appears to be little correlation of the measured value of $\kappa_0$ with the three different cells used. Cell 204F has the lowest total pressure and a 5\% smaller value for $\kappa_0$ than the other two cells. As discussed in Sec.\ \ref{theory}, we would not expect this to be due to $\kappa_1$ in Eq.~(\ref{eq:kappaAX}), as there is enough total gas pressure in all of these cells to assure that the van der Waals molecules are in the short-lifetime limit \cite{happer1984polarization}. Any temperature dependence would appear to be weak at best; this is not unexpected due to the steep core wall of the RbXe van der Waals potential and the strong dependence of the contact interaction on the distance of closest approach. Looking at the overall scatter in the data, the 1.5\% relative uncertainty in the weighted average may be a bit optimistic. If so, the source of any systematic error is likely to come from the NMR ratio measurement, since the frequency-shift measurements have better SNR and involve less post-processing to arrive at the $^{129}$Xe-$^3$He signal ratio. Despite our careful efforts to calibrate the NMR equipment, drifts in sensitivity during the measurements due to thermally sensitive components or to inhomogeneities in the transverse excitation field cannot be completely ruled out.

In Sec.\ \ref{NMRacq}, we assumed that no additional $^{129}$Xe nuclear polarization is generated by SEOP during the $\approx 5$~s interval over which $^{129}$Xe NMR signals are rapidly acquired and averaged. To justify this assumption, we use the known binary spin-exchange rate coefficient $k_{\rm se}=2.2\times 10^{-16}$~cm$^3$/s for Rb-$^{129}$Xe \cite{Jau2003} and the Rb vapor pressure curve due to Killian \cite{Killian1928} to estimate the spin-exchange time $\gamma_{\rm se}^{-1}$ to be in range $13-170$~s for the temperature range $T=175-125\ ^{\circ}$C. The wall-relaxation time $\Gamma_w^{-1}$ in these cells is a few minutes or more, so most of our measurements (e.g., the one shown in Fig.~2) well satisfy $\tau_up\gg 5$~s, where the ``spin-up" time $\tau_{\rm up}=(\gamma_{rm se}+\Gamma_w)^{-1}$. Moreover, since the flip angle $\theta_{\rm Xe}\ll 1$~rad, many if not most of the acquisitions over the 5-s interval occur near the steady-state polarization maximum, where the slope of the spin-exchange transient is smallest and the corresponding polarization recovery is weakest.
 
We further assumed for both the EPR and NMR measurements that the nuclear magnetization is uniform throughout the cell for both nuclear species at the time any measurement is initiated. The relevant diffusion coefficients \cite{Barbe1974,Acosta2006} are such that $^3$He atoms traverse these cells in a characteristic time $\tau_d$ of a few seconds at most and that Xe atoms do so in no more than about 10-15~s. Since the condition $\tau_d\ll\tau_{\rm up}$ always holds for $^3$He, there is never a question that the $^3$He magnetization is uniform. The same condition usually also holds for $^{129}$Xe, although at our highest measurement temperatures it may only be marginally satisfied. Nevertheless, if the cell is approximately uniformly illuminated and the initial EPR frequency measurement (prior to polarization destruction) is made after several times $\tau_{\rm up}$, any residual nuclear polarization gradient in the cell should be very small. Thus, for the few measurements made at the highest temperatures, there may have been some very small amount of recovery during the $^{129}$Xe NMR acquisition, as well as a small $^{129}$Xe polarization gradient, but there are no obvious trends in the data at high temperatures to suggest that we cannot ignore both of these effects.

Another systematic effect we considered was cell asphericity: both the slightly oblate or prolate shape of the nominally spherical cell and the ``pull-off," the roughly cylindrical sub-volume protruding radially out from the main cell volume. The pull-off is formed as a by-product of flame-sealing the cell away from a glass manifold after it has been filled with a few tens of milligrams of alkali-metal and the constituent gases. A perfectly spherical distribution of a uniformly magnetized material produces zero through-space field in the interior of the sphere. Asphericity produces a net non-zero through-space field which would also be probed by the EPR laser, producing a frequency shift that cannot be distinguished from that due to the Fermi-contact interaction. We find that the effects of asphericity contribute at no more than about the 0.4\% to our results, significantly smaller than our quoted statistical error; details are discussed in the Appendix.
 
With appropriate rearrangement of Eqs.~(\ref{eq:EPRshift}) and (\ref{eq:magdef}), our result can be used to calibrate the measured EPR frequency shift to the absolute $^{129}$Xe polarization $P_{\rm Xe} = \langle K_z\rangle/K$. This relationship also shows that the largest measured EPR shifts in our cells, combined with knowledge of the Xe density, sets a lower bound for $(\kappa_0)_{\rm RbXe}$. In the case of these measurements, a maximum shift of $5.0$ kHz was recorded for cell 204F, yielding a $^{129}$Xe polarization of 87\% using our measured value of $(\kappa_0)_{\rm RbXe}$; alternatively, by assuming 100\% polarization, the measurement of that same shift sets a lower bound on $(\kappa_0)_{\rm RbXe}$ of 452.

To put our measurement into context with those preceding it, we first note that the result agrees well statistically with the results of Ma, et. al.\ \cite{ma2011collisional}, who measured $(\kappa_0)_{\rm RbXe}= 493\pm 30$ using an entirely different experimental method (comparing the NMR frequency shifts for $^3$He and $^{129}$Xe in the presence of a polarized $^{87}$Rb vapor), although they similarly relied on the previous $\kappa_0$ result for $^{87}$Rb-$^3$He \cite{romalis1998accurate}. The present result is somewhat smaller than the recent theoretical/computational results of Hanni, et al.\ \cite{hanni2017electron} who determined $(\kappa_0)_{\rm RbXe}= 588\pm 50$; but the discrepancy is not alarming considering the uncertainties. Earlier estimates by Walker \cite{Walker1989} and measurements by Schaefer, et al.\ \cite{schaefer1989frequency} were in the range of 650 to 750, but had much larger uncertainties, and so are also not inconsistent with our result. Finally, note that this work may be relevant to SEOP of radon for measurements of the permanent electric-dipole moment (EDM) \cite{Chupp2019RMP,Tardiff2014,Kitano1988}; radon is a strong candidate for observing a non-zero EDM and/or setting the best experimental limit on its value. The enhancement factor $\kappa_0$ is predicted to be as large as 1400 for the Rb-Rn system \cite{Walker1989}; the effects of the associated frequency shifts may thus play a significant role in these precision measurements.

\begin{acknowledgments}
The authors thank University of Utah glassblower K.~Teaford for cell and sample chamber manufacturing and acknowledge the U.S. National Science Foundation grant PHY-0953225 for support of this work.
\end{acknowledgments}

\appendix*
\section{Effects of cell asphericity}
Here we estimate the effects of having imperfectly spherical cells. Since $(\kappa_0)_{\rm RbHe}\approx 5$ is on the order of unity and much smaller than $(\kappa_0)_{\rm RbXe}$, the non-zero through-space field in non-spherical geometries can be an important systematic effect in deducing the correct EPR frequency shift and relating it to $\kappa_0$. Since the effect produces a much larger fractional change in $(\kappa_0)_{\rm RbHe}$, we treat any fractional systematic error in the measurement of the EPR shift due to $^3$He as the fractional error in the ratio measurement. We start with (in Gaussian units)

\begin{equation}
\label{eq:CGSH}
{\mathbf B}={\mathbf H}+4\pi{\mathbf M}
\end{equation}

\noindent and recall that the $\delta$-function contribution to the field in any magnetized region is given by ${\mathbf B}_{\delta} = \frac{8\pi}{3}{\mathbf M}$. We consider two possible sources of these effects: global asphericity of the cell and the cell pull-off.

\subsection{Modeling the pull-off}
Figure~\ref{fig:IC} shows an idealized drawing of a spherical vapor cell with a cylindrical pull-off with relevant dimensions given in units of the cell radius $R$. In our experiments the pull-off was oriented toward the floor, and the applied field was horizontal, as depicted in the figure. The total magnetic moment inside the cylindrical pull-off is $\mu_{\rm cyl}=\pi a^2bR^3M_0$, where $M_0$ is the uniform magnetization of the cell, $aR$ is the pull-off radius, and $bR$ is the pull-off length. We consider the through-space field at the center of the cell, although the narrow (relative to $R$) probe laser beam nominally runs along a line through the cell center; the field will thus be a little smaller as one approaches the entrance and exit points of the probe beam.

As a first crude approximation, we assume that the total magnetic moment of the pull-off is a point dipole located in the middle of the pull-off, i.e., that it is $R+\frac{1}{2}bR$ away from the center of the cell. This will generate a field $B_{\rm pull}$ at the cell center given by

\begin{equation}
\label{eq:Bpull1}
B_{\rm pull}=-\frac{\mu_{\rm cyl}}{R^3(1+\frac{1}{2}b)^3}=-\frac{\pi a^2b}{(1+\frac{1}{2}b)^3}M_0,
\end{equation}

\noindent where the negative sign indicates that ${\mathbf B}_{\rm pull}$ is antiparallel to the magnetization ${\mathbf M_0}$. For typical values in the cells used in this work, $a=0.2$, $b=1.2$, $B_{\rm pull} = -0.037M_0$.

A better approximation assumes a uniform line of dipoles occupies the pull-off with linear magnetization density $\pi a^2R^2M_0$; we then integrate the pull-off field from $z=R$ to $z=(1+b)R$:

\begin{equation}
\label{eq:Bpull2}
B_{\rm pull}=-\int_R^{R(1+b)}\frac{\pi a^2R^2}{z^3}M_0dz=\frac{\pi a^2}{2}\Bigg[\frac{1}{(1+b)^2}-1\Bigg]M_0.
\end{equation}

\noindent For the same values of $a$ and $b$, we now obtain $B_{\rm pull} = -0.050M_0$. As expected, the magnitude of this estimate is larger than the first one due to the $1/r^3$ drop-off of the dipolar field. If we divide our result by the contact field $\frac{8\pi}{3}(\kappa_0)_{\rm RbHe}M_0\approx 40M_0$, it translates directly to a fractional error in our ratio measurement, meaning that the pull-off might systematically depress the measured value of $(\kappa_0)_{\rm RbXe}$ at about the 0.1\% level, a small change compared to our quoted statistical error.

\subsection{Modeling global asphericity}

The through-space field is uniform and non-zero for uniformly magnetized ellipsoids. We model the global cell asphericity as a slight difference in major axis along one of the coordinate axes---in other words, as an oblate or prolate spheroid---and calculate the slight deviation of the interior through-space field from zero. The general calculation is not trivial; a good reference is due to J.A. Osborn \cite{Osborn1945}. The demagnetizing tensor ${\cal N}$ is defined by

\begin{equation}
\label{eq:demagMKS}
{\mathbf H}_{\rm d}=-4\pi{\cal N}\otimes{\mathbf M},
\end{equation}

\noindent where ${\mathbf H}_{\rm d}$ is the value of the magnetic field (H-field) that produces the correct magnetic induction (B-field) in a given magnetized geometry. In highly symmetric situations with uniform magnetization, the tensor is diagonalized to values along the three principal axes, $N_x$, $N_y$, and $N_z$. For a sphere magnetized along the z-axis, $N_z=1/3$, yielding the result that the B-field inside the sphere is due entirely to the $\delta$-function contribution.

Osborn \cite{Osborn1945} calculated the principal demagnetizing factors for a general ellipsoid with principal axes $a$, $b$, and $c$ along the x-, y-, and z-coordinate axes, respectively. We reproduce here his results for the special cases of prolate and oblate spheroids with the symmetry axis in the z-direction. For a prolate spheroid ($c>a$, $a=b$), with $m\equiv c/a$:

\begin{equation}
\label{eq:prolatez}
N^{\rm p}_z = \frac{1}{m^2-1}\Bigg[\frac{m}{2\sqrt{m^2-1}}\log\Big(\frac{m+\sqrt{m^2-1}}{m-\sqrt{m^2-1}}\Big)-1\Bigg],
\end{equation}

\noindent whereas for an oblate spheroid ($c<a$, $a=b$, $m\equiv c/a$),

\begin{equation}
\label{eq:prolatez}
N^{\rm o}_z = \frac{1}{1-m^2}\Bigg[1-\frac{m}{\sqrt{1-m^2}}\arcsin{\sqrt{1-m^2}}\Bigg].
\end{equation}

\noindent We consider a 5\% difference in major axes: for $c/a = 1.05$ we obtain $N_z^{\rm p}=0.32042$ and for $a/c = 1.05$ we obtain $N_z^{\rm o}=0.34643$. The resulting non-zero through-space B-field produced in our cells by these demagnetizing factors is:

\begin{equation}
\label{eq:ts}
B_z^{\rm ts}=4\pi(\frac{1}{3}-N_z)M_0.
\end{equation}

\noindent Equation~(\ref{eq:ts}) yields $B_z^{\rm ts}=\pm4\pi(0.0129)M_0$ for the prolate (oblate) spheroid. If we again divide these results by the contact field $\frac{8\pi}{3}(\kappa_0)_{\rm RbHe}M_0\approx 40M_0$, the corresponding systematic contribution to the measured value of $(\kappa_0)_{\rm RbXe}$ occurs at about the 0.4\% level. We also note that whereas any effect due to the pull-off  acts to depress the value of $(\kappa_0)_{\rm RbXe}$, the sign of the effect for global asphericity depends on whether the cell is prolate or oblate.

\pagebreak

\begin{table}
\caption{\label{tab:tab1}
Gas composition (in torr at 20~$^{\circ}$C) of the three cells used in this work. Cell volume is shown in parenthesis below the cell designation.}
\begin{ruledtabular}
\begin{tabular}{lllllllll }
&{Cell 203C}&{Cell 204F}&{Cell 205B}\\
&{($7.75~\mathrm{cm}^3$)}&{($7.61~\mathrm{cm}^3$)}&{($7.1~\mathrm{cm}^3$)}\\
\\[-0.9em]
\hline
\\[-0.6em]
Helium  & 2185 $\pm$ 85	& 1639 $\pm$ 65		& 2103 $\pm$ 80		\\
Xenon   & 41 $\pm$ 2		& 22 $\pm$ 1 			& 18 $\pm$ 1			\\
Nitrogen& 45 $\pm$ 2		& 34 $\pm$ 1			& 43 $\pm$ 2			\\
Total   & 2271 $\pm$ 85	& 1695 $\pm$ 65		& 2164 $\pm$ 80		\\
\end{tabular}
\end{ruledtabular}
\end{table}

\begin{table}
\caption{\label{tab:tab2}
Number of measurements (in bold) shown with the uncertainty-weighted average (with the standard deviation of the mean) of $(\kappa_0)_{\rm RbXe}$ for each cell at each temperature measured.}
\begin{ruledtabular}
\begin{tabular}{cccccc}
							&{$125~^{\circ}$C}&{$150~^{\circ}$C}&{$165~^{\circ}$C}&{$175~^{\circ}$C}&{Average}\\
\hline\\[-0.8em]
\multirow{2}{3em}{Cell 203C}	&$\mathbf{3}$&$\mathbf{10}$&	&$\mathbf{4}$&$\mathbf{17}$ \\
							& $596\pm21$ & $519\pm13$ & ... & $537\pm49$ & $532\pm14$\\\\[-0.8em]
\hline
\\[-0.8em]
\multirow{2}{3em}{Cell 204F}	&$\mathbf{1}$&$\mathbf{12}$&$\mathbf{12}$&	&$\mathbf{25}$ \\
							&$585\pm53$&$490\pm15$&$523\pm21$&...&$501\pm12$\\\\[-0.8em]
\hline
\\[-0.8em]
\multirow{2}{3em}{Cell 205B}	&$\mathbf{5}$& &$\mathbf{1}$&	&$\mathbf{6}$	\\
							&$527\pm26$&...&$520\pm29$&...&$526\pm22$\\\\[-0.8em]
\hline\\[-0.8em]
\multirow{2}{3em}{Average}		&$\mathbf{9}$&$\mathbf{22}$&$\mathbf{13}$&$\mathbf{4}$&$\mathbf{48}$\\
							&$551\pm20$&$505\pm10$&$523\pm19$&$537\pm49$&$518\pm8$\\
\end{tabular}
\end{ruledtabular}
\end{table}

\begin{figure*}
\includegraphics[width=\textwidth]{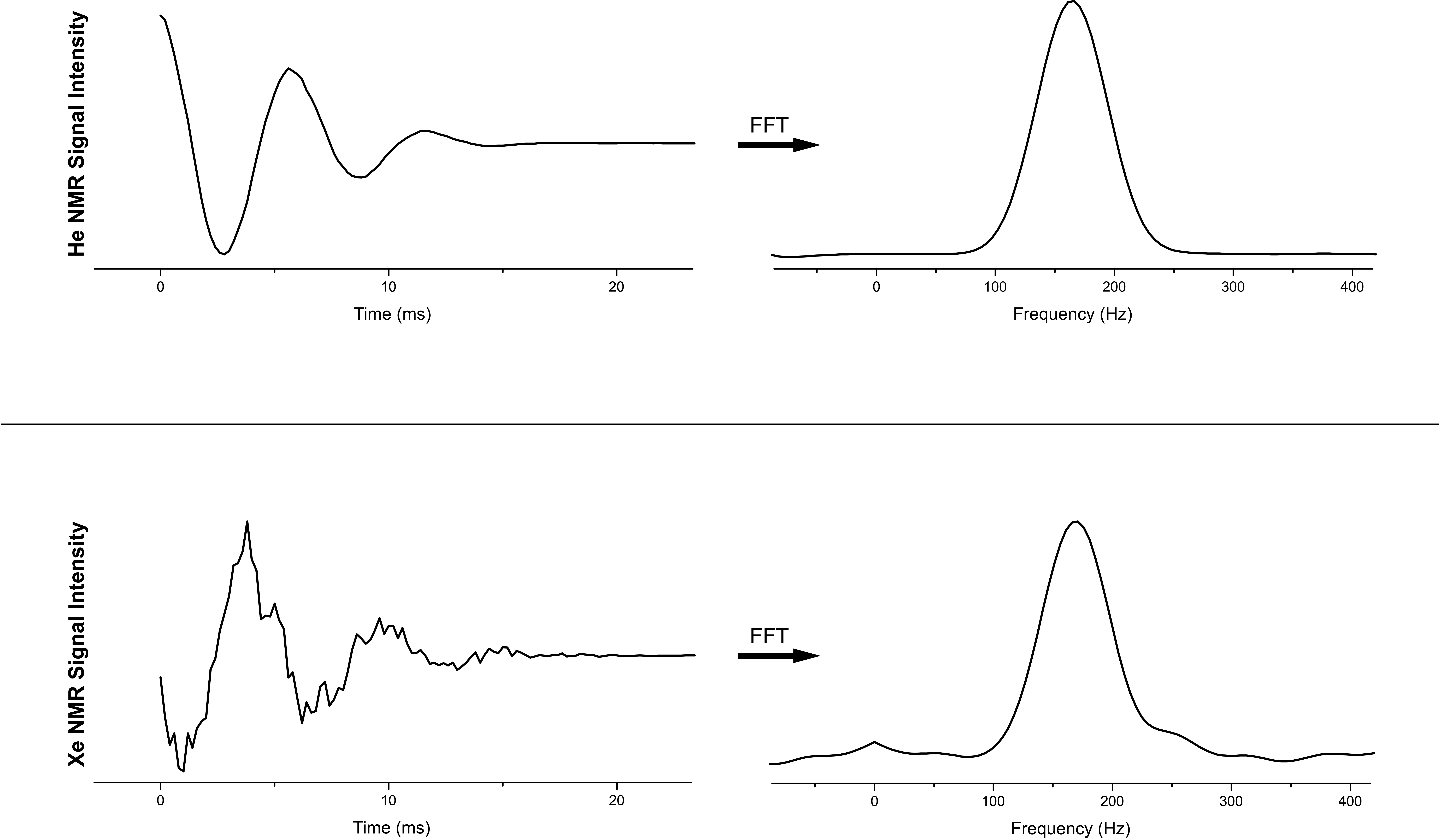}
\caption{\label{fig:fig1}$^3$He (top) and $^{129}$Xe (bottom) NMR signals. The left side shows the time-domain free-induction decay (FID) and the right side shows the corresponding Fourier transform relative to the carrier frequency. Signals were acquired at the same frequency (31.25~kHz) with the same RF power and excitation pulse length.
}
\end{figure*}

\begin{figure*}
\includegraphics[width=\textwidth]{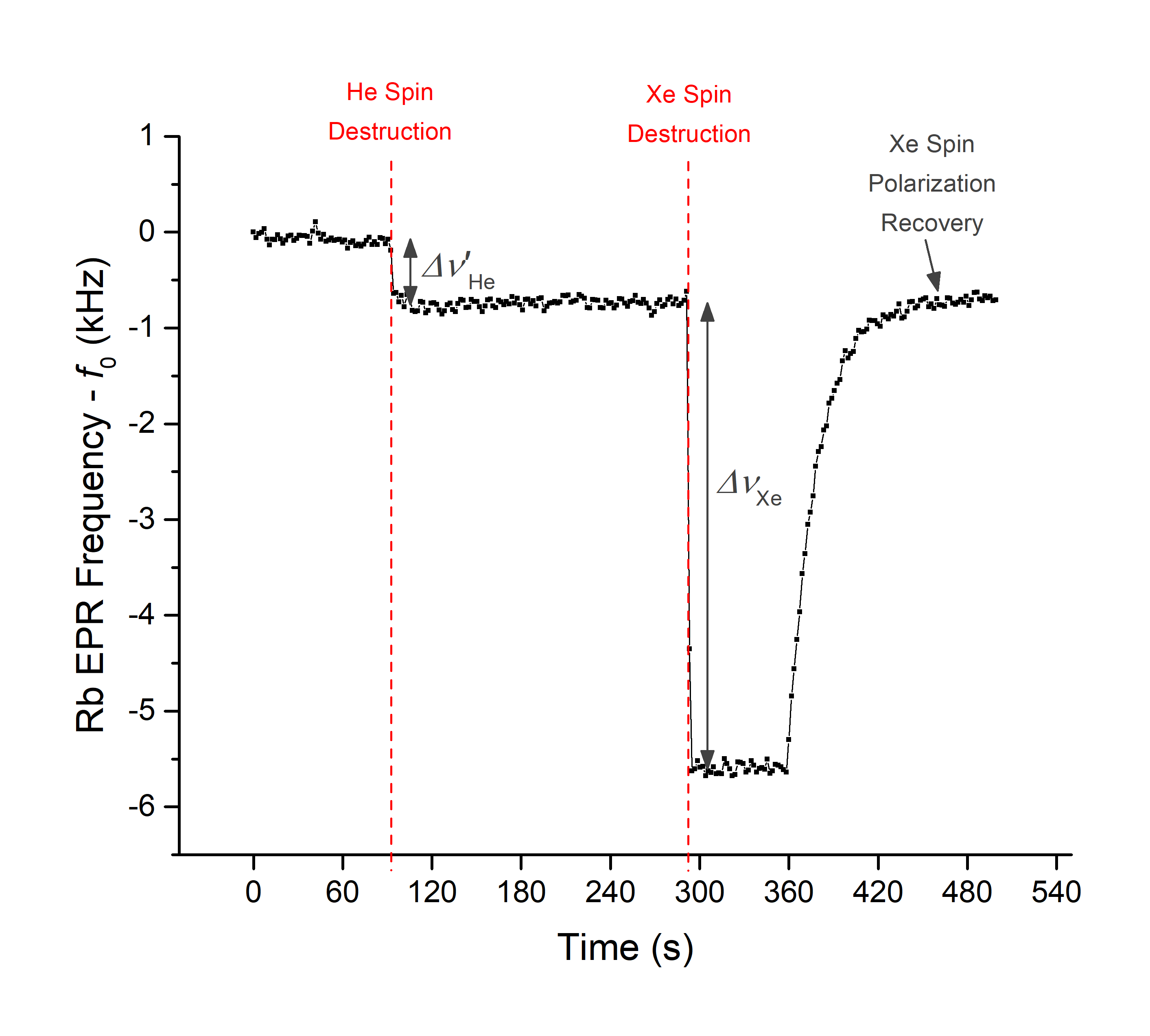}
\caption{\label{fig:fig2}$^{87}$Rb EPR frequency shift as a function of time, measured relative to the initial recorded frequency $f_0 = 18.6$~MHz in cell 204F at 150~$^{\circ}$C. At $t=0$ both $^3$He and $^{129}$Xe were fully polarized. At $\approx 90$~s, the $^3$He magnetization was selectively destroyed with a comb of resonant (85.6~kHz) NMR pulses causing a frequency shift $\Delta\nu_{\rm He}^{\prime}$. At $\approx 290$~s, the $^{129}$Xe magnetization was similarly destroyed, yielding a frequency shift $\Delta\nu_{\rm Xe}$. The $^{129}$Xe magnetization recovered in a few minutes to its original value; the $^3$He magnetization showed no appreciable recovery throughout the experimental run.}
\end{figure*}

\begin{figure}
\includegraphics[scale=0.18]{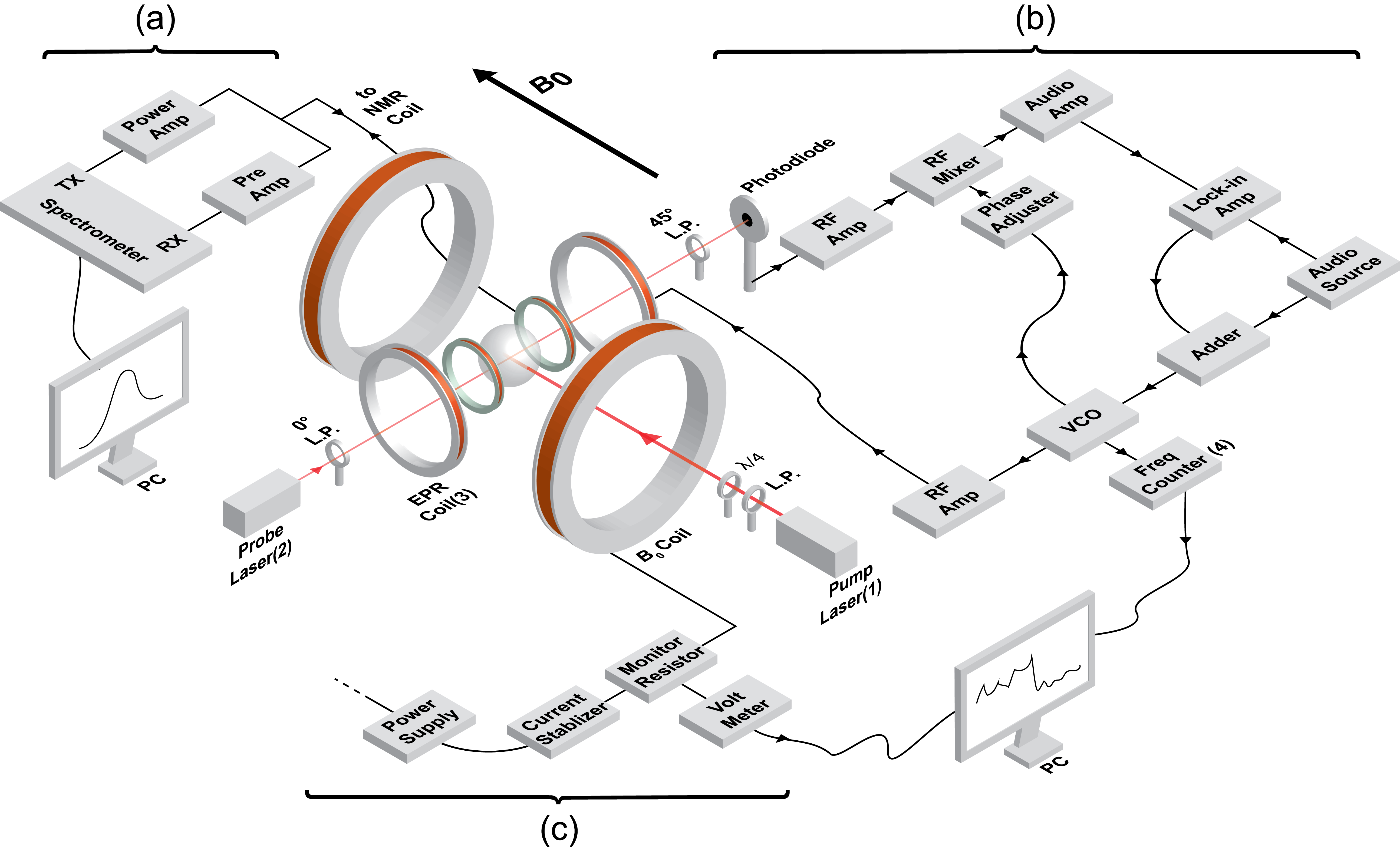}
\caption{\label{fig:fig3}Schematic of experimental apparatus; model/manufacturer designations not noted here appear in the text. (a) The NMR spectrometer transmits excitation pulses to and receives the subsequent FID signal from a tuned coil placed near the cell. (b) Optically detected EPR consists of (1) The 795~nm pump laser with 10-12~W of power narrowed to $\sim$0.3 nm. (2) An 80-mW probe laser detuned by $\approx 1$~nm that probes the Rb magnetization via Faraday rotation; the transmitted light (intensity modulated at the $\approx 19$~MHz EPR frequency) is focused onto a fast photodiode. (3) A two-turn tuned EPR coil, located coaxially but outside the NMR coil, is driven by a voltage-controlled oscillator (VCO) through a ZHL-32A RF amplifier (Mini-Circuits). The ZAD-1 RF mixer (Mini-Circuits) homodynes the photodiode signal with the VCO output; the difference signal is fed to a model 186A lock-in amplifier (Princeton Applied Research) referenced to a 100-Hz sine wave that also modulates the VCO frequency with an amplitude much smaller than the transition linewidth. The derivative (error) signal at the lock-in output locks the frequency to the peak of the $^{87}$Rb hyperfine resonance. (4) A precision counter records the output frequency from the VCO and sends it to a PC-type computer running a LabVIEW program for display and analysis. (c) The applied magnetic field (26.5~G) is generated by a 60-cm-dia.\ Helmholtz pair, the current stabilizer described in Fig.~5, and a 0.10-$\Omega$ monitor resistor. The voltage across the monitor resistor is used to correct the residual long-term current drift.
}
\end{figure}

\begin{figure}
\includegraphics[width=\textwidth]{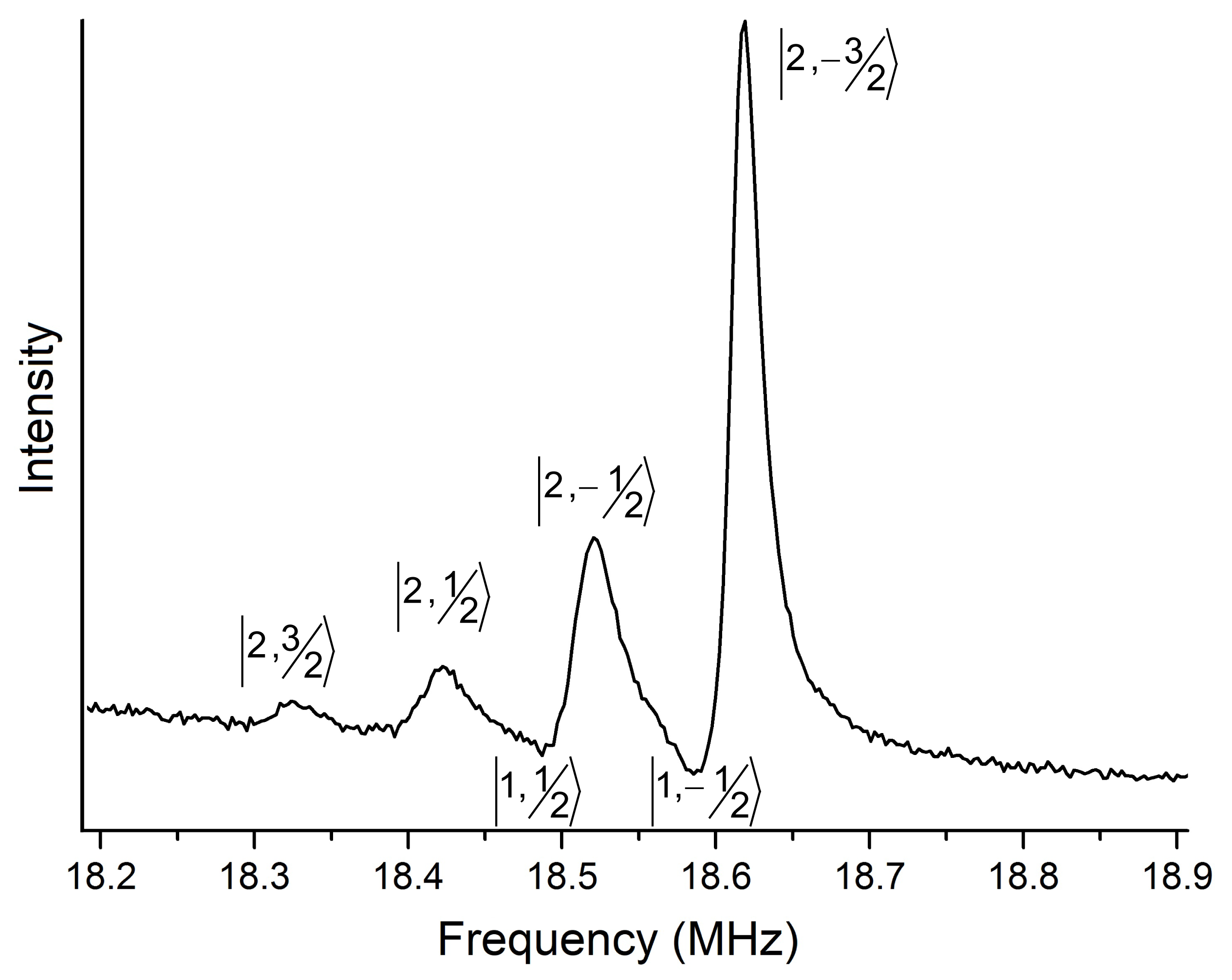}
\caption{\label{fig:fig4}Optically-detected $^{87}$Rb EPR hyperfine spectrum under optical pumping conditions at 26.5~G from cell 204A; peaks are labeled by $\vert F,\overline{m}_F\rangle$. A voltage-controled oscillator (VCO) swept with a wide voltage ramp across the resonances generates the full hyperfine spectrum. Here, most of the intensity resides in the $F=2$, $\overline{m}_F=-3/2$ transition, because the $\sigma^-$ pumping light drives population towards the $\vert 2,-2\rangle$ state. The $F=1$ transitions are barely discernable (180$^{\circ}$ out of phase with $F=2$) to either side of the $\vert 2,-\frac{1}{2}\rangle$ peak.}
\end{figure}

\begin{figure}
\includegraphics[width=\textwidth]{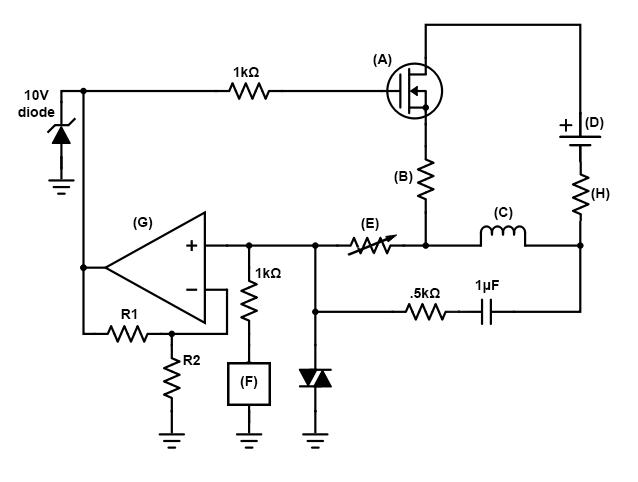}
\caption{\label{fig:fig5}Current-stabilizing circuit design: a power MOSFET (A) (IRL2910) is placed in series with a high-power shunt resistor (B) (Vishay-Dale 250 W, .5 $\Omega$, 100 ppm/$^{\circ}$C) and a high-precision variable resistor (E) (Vishay Accutrim 1240, 0-500 $\Omega$, 10 ppm/$^{\circ}$C). The shunt resistor is submerged in mineral oil to increase thermal mass and minimize temperature change due to air flow. The voltage across the resistors is compared to a stable voltage reference (F) (MAX6341) and fed into the gate terminal of the transistor through a non-inverting OP-27 operational amplifier (G) in a negative-feedback mode. The gate voltage controls the drain-source current which is approximately the same as the one flowing through the coils (C). The 1-$\mu$F capacitor is used to damp any current oscillation from the coils. The crossed diodes are used to protect the op-amp. (H) The 0.10-$\Omega$ sensing resistor (H) is used to correct the current drifts.}
\end{figure}

\begin{figure}
\includegraphics[width=\textwidth]{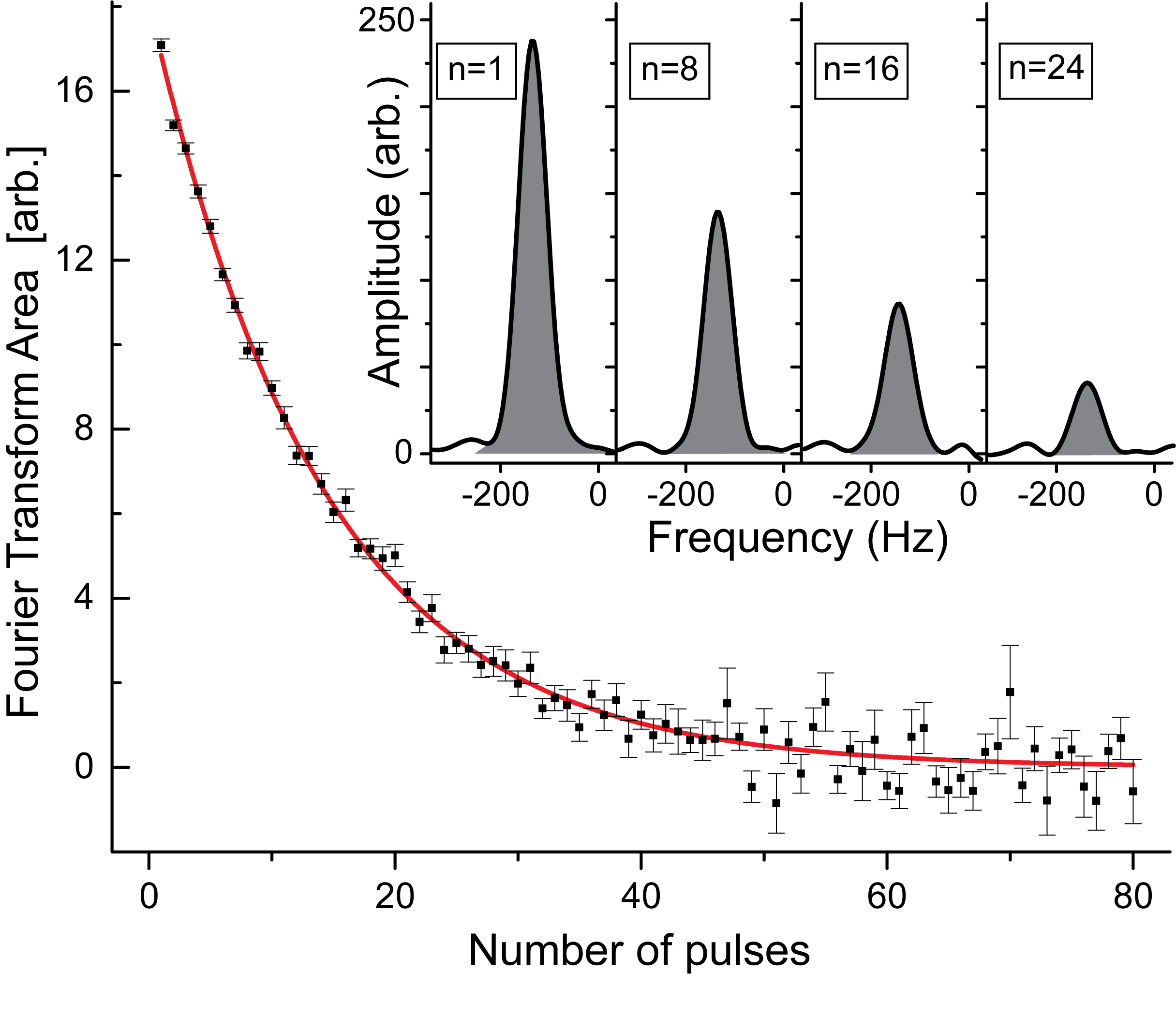}
\caption{\label{fig:fig6}Signal intensity as a function of pulse number for eighty consecutive 80-$\mu$s pulses at the $^{3}$He Larmor frequency, 31.25 kHz; the rate is one pulse every 55~ms. The $^{3}$He FID after each pulse is collected, digitized and Fourier transformed. The peak area is plotted vs.\ the number $n$ of pulses and fit to an exponential decay (red line); see Eq.~(\ref{eq:decay}).
}
\end{figure}

\begin{figure}
\includegraphics[width=\textwidth]{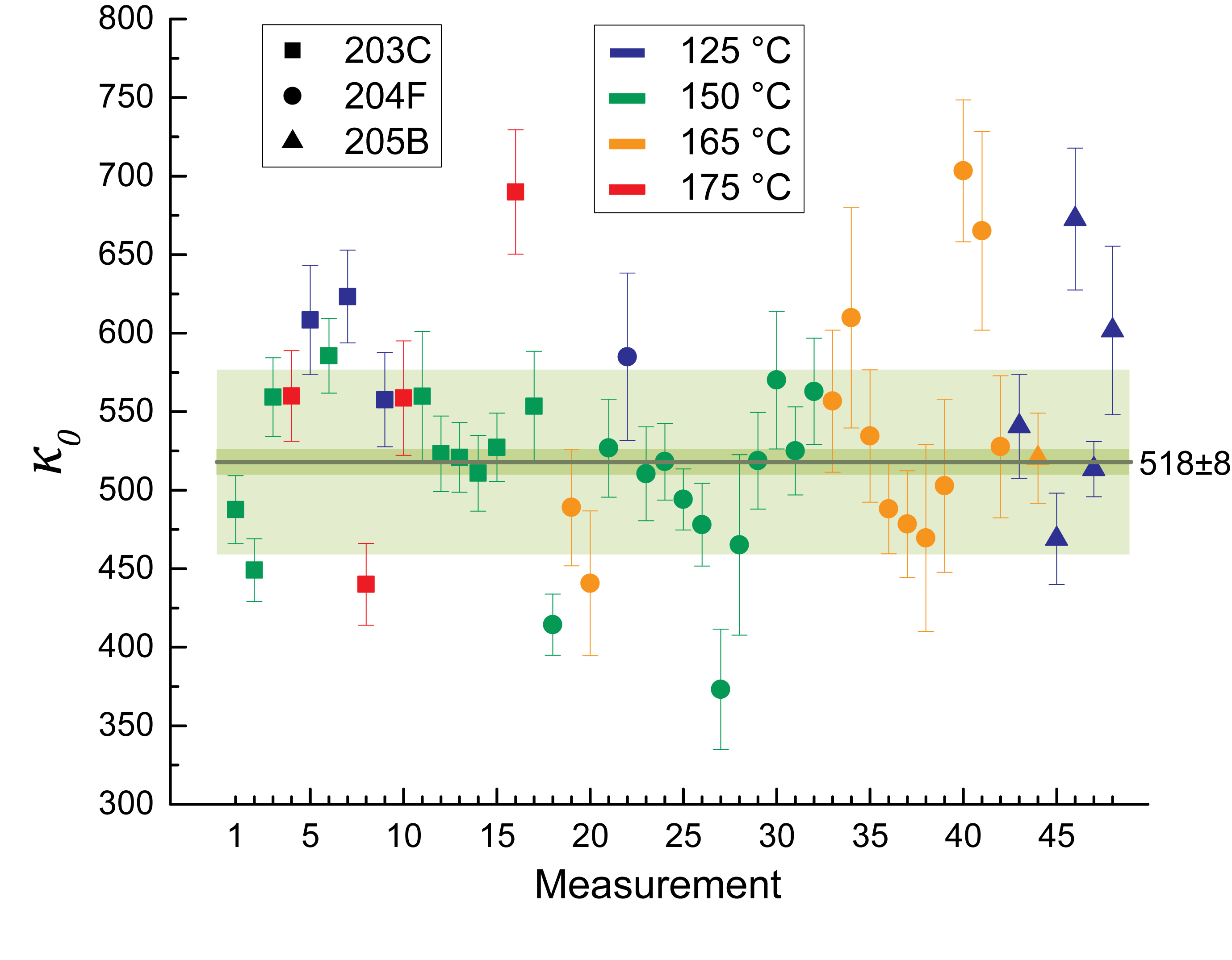}
\caption{\label{fig:fig7}Calculated values of $\kappa_0$ with the corresponding error bars from 48 measurements of three cells at four temperatures. The uncertainty-weighted average of $\kappa_0$ with its standard error is shown (dark green band) along with the weighted standard deviation ($\pm 59$) for the whole set of data (light green band).}
\end{figure}

\begin{figure}
\includegraphics[width=2in]{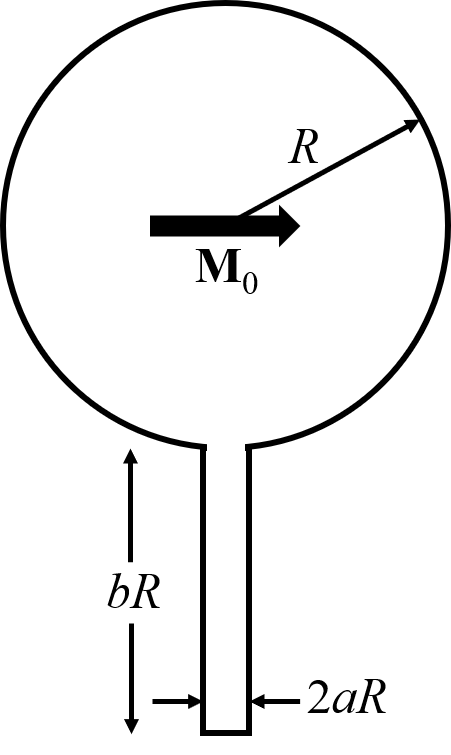}
\caption{\label{fig:IC}Vapor cell idealized as a perfect sphere connected to a cylinder that models the cell pull-off.}
\end{figure}

\end{document}